\documentclass[bibyear]{aa}

\usepackage{graphicx}
\usepackage{amsmath}
\usepackage{bm}
\usepackage{float}
\usepackage{caption}

\newcommand\kms{\ifmmode{\rm km\thinspace s^{-1}}\else km\thinspace s$^{-1}$\fi}

\begin{document}

\title{ Limb-Darkening Coefficients for the 4-Term and Power-2 Laws
  for the JWST Space Mission, Adopting PHOENIX Spherical Models at
  High Resolution
  \thanks{Tables 2 and 3 are only available at the CDS via anonymous ftp
  to cdsarc.u-strasbg.fr (130.79.128.5) or via
  http://cdsarc.u-strasbg.fr/viz-bin/cat/J/A+A/XXX/XXX }}

\subtitle{NIRCam, NIRISS and NIRSpec passbands}

\titlerunning{Limb-darkening coefficients for the 4-term and power-2 laws}

\author{A.\ Claret~\inst{1,2}  \and  P.\ H.\ Hauschildt\inst{3}   \and G.\ Torres \inst{4}}

\offprints{A.\ Claret, e-mail:claret@iaa.es}

\institute{Instituto de Astrof\'{\i}sica de Andaluc\'{\i}a, CSIC, Apartado 3004, 18080 Granada, Spain
      \and Dept.\ F\'{\i}sica Te\'{o}rica y del Cosmos, Universidad de Granada, Campus de Fuentenueva s/n,  
      10871, Granada, Spain
      \and {Hamburger Sternwarte, Gojenbergsweg 112, 21029 Hamburg, Germany}
      \and {Center for Astrophysics $\vert$ Harvard \& Smithsonian, 60 Garden St., Cambridge, MA 02138, USA }
}

\date{Received; accepted; }

\abstract
    {}
    {Modeling observations of transiting exoplanets or close binary
    systems by comparing the observations with theoretical light
    curves requires precise knowledge of the distribution of specific
    intensities across the stellar disk.  We aim to facilitate this
    type of research by providing extensive tabulations of
    limb-darkening coefficients for 11 frequently used near- and
    mid-infrared passbands on the NIRCam, NIRISS, and NIRSpec
    instruments installed on board the James Webb Space Telescope.}
    {The calculation of the limb-darkening coefficients was based on
    spherically symmetric atmosphere models from the PHOENIX series,
    with high spectral resolution (approximately 10$^{6}$
    wavelengths), and covering the wavelength range 0.1--6.0~$\mu$m.  The models
    were computed for solar composition, and a microturbulent velocity
    of 1.0~\kms. We adopted two of the more accurate parametrizations
    for the coefficients: the 4-term law, and the power-2 law. We
    applied the Levenberg–Marquardt least-squares minimization method,
    with a strategy to determine the critical value $\mu_{\rm crit}$
    of the cosine of the viewing angle near the limb that is designed
    to improve numerical accuracy.}
    {The limb-darkening coefficients were derived based on a total of
    306 atmosphere models covering an effective temperature range of
    2400--7800~K, and a $\log g$ interval between 3.0 and 5.5.  We
    discuss the quality of the fits to the specific intensities
    provided by the power-2 and 4-term laws, as well as by the often
    used quadratic law. Based on a comparison, we recommend the use of
    the 4-term or power-2 laws, in that order of preference.}
    {}


\maketitle 

\section{Introduction}

Limb-darkening is an important phenomenon in several areas of
Astrophysics research, including transiting extrasolar planets,
eclipsing binary stars, spectroscopy, stellar microlensing events,
interferometry, exozodiacal dust, line profiles of rotating stars, and
many others.  The correct treatment of this effect can be critical to
the accuracy of the results, particularly in the present era of very
high-precision observations made from space.

One of the first observations of the limb-darkening phenomenon was
carried out centuries ago by Bouguer (1760). It was not until the
beginning of the 20th Century that theoretical research into the
limb-darkening phenomenon began, led by Schwarzschild (1906). This
author introduced a linear law to describe the distribution of
specific intensities as a function of a parameter $\mu = \cos\gamma$,
where $\gamma$ is the angle between the observer's line of sight and
the surface normal (see also Milne 1921). That well-known linear
equation is written as
\begin{equation}
  \frac{I(\mu)}{I(\mu=1)}=  1 - u(1 - \mu),
\end{equation} 
where $I(\mu)$ is the specific intensity as a function of position on
the stellar disc, and $u$ is the linear limb-darkening coefficient
(LDC). Some four decades later, Chandrasekhar (1944) and Placzek
(1947) worked out some results concerning limb-darkening for the grey
atmosphere case.

A number of other laws with an extra term have been proposed
since. For example, Kopal (1950) introduced the quadratic law
\begin{equation}
  \label{eq:quadratic}
  \frac{I(\mu)}{I(\mu=1)}=  1 - a(1 - \mu) - b(1-\mu)^2,
\end{equation}
which has been quite popular in recent years.  A decade later,
Van't Veer (1960) proposed the following equation for the grey case:
\begin{equation}
  \frac{I(\mu)}{I(\mu=1)}=  1 - b_1(1 - \mu) - b_3(1-\mu)^3,    
\end{equation}  
and  Klinglesmith \& Sobieski (1970) introduced the logarithmic law 
\begin{equation}
  \frac{I(\mu)}{I(\mu=1)}=  1 - e(1 - \mu) - f~\mu~\log(\mu). 
\end{equation} 
Another bi-parametric equation, the square-root law by
D{\'{\i}}az-Cordov\'es \& Gim\'enez (1992), has the expression
\begin{equation}
  \label{eq:sqrt}
  \frac{I(\mu)}{I(\mu=1)}=  1 - c(1 - \mu) - d(1-\sqrt{\mu}). 
\end{equation} 
Compared to the linear formula, these bi-parametric laws yielded
improved agreement with the specific intensities provided by stellar
atmosphere models, particularly in the case of Eq.(\ref{eq:sqrt}) for
stars with high effective temperatures.  In the above equations, $a$,
$b_1$, $e$ and $c$ are the linear LDCs, while $b$ is the quadratic
LDC, $b_3$ the cubic LDC, $f$ is the logarithmic LDC, and $d$ the
square-root LDC.
 
The best fit to model specific intensities provided by a bi-parametric
law is due to Hestroffer (1997), who introduced a prescription that
involves a power of $\mu$, the so-called power-2 law. For a detailed
discussion of the properties and advantages of this law, see Claret \&
Southworth (2022), their Sections~3 and 4.  The power-2 law is
expressed as
\begin{equation}
  \frac{I(\mu)}{I(\mu=1)}=  1 - g(1 - \mu^h),
\end{equation} 
where $g$ and $h$ are  the corresponding LDCs.  

More recently, Claret (2000) proposed a 4-term non-linear law written
as
\begin{eqnarray}
  \frac{I(\mu)}{ I(1)} = 1 - \sum_{k=1}^{4} {a_k} (1 - \mu^{\frac{k}{2}}), 
\end{eqnarray}
in which $a_k$ are the associated LDCs. So far, this 4-term
limb-darkening prescription has achieved the best fits to the specific
intensities from stellar atmosphere models, for both those using the
plane-parallel and spherical geometry.

Finally, Sing et al.\ (2009) have advocated for an abbreviated form of
the 4-term law given by
\begin{equation}
  \frac{I(\mu)}{I(\mu=1)}= 1 - a_2(1 - \mu) - a_3(1 - \mu^{3/2}) -
  a_4(1 - \mu^2),
\end{equation} 
in which the $\mu^{1/2}$ term is omitted.  Those authors pointed out
that ``The $\mu^{1/2}$ term from the four parameter non-linear law
mainly affects the intensity distribution at small $\mu$ values, and
is not needed when the intensity at the limb varies approximately
linearly at small $\mu$ values''. We note, however, that not all
atmosphere models show a quasi-linear drop in intensity near the limb,
to a degree that warrants the omission of that term. Models with
non-negligible curvature include those based on spherical geometry, or
those for hot stars (see our comment above regarding
Eq.(\ref{eq:sqrt}), which does include the $\mu^{1/2}$ term).

The advent of space missions has come with tighter requirements on
precision in dealing with the phenomenon of limb darkening.  Until
recently, this was a concern mostly at optical wavelengths, where the
effect is more noticeable.  However, the launch of the James Webb
Space Telescope (JWST), at the end of 2021, represents a significant
advancement, among others, in our ability to explore the properties of
exoplanets in the near- and mid-infrared, a region of the spectrum
rich in new information. JWST has opened the door to detailed chemical
studies of the complex atmospheres and compositions of these objects,
through the analysis of light curves and/or transmission spectroscopy
of unprecedented quality. These observations are now being gathered
with the Near Infrared Camera (NIRCam), the Near Infrared Imager and
Slitless Spectrograph (NIRISS), and the Near Infrared Spectrograph
(NIRSpec) on board the spacecraft, as well as with the Mid Infrared
Instrument (MIRI) at even longer wavelengths.  To take full advantage
of these new observations, proper treatment of limb-darkening and
other effects has become more important.  The main goal of this work,
therefore, is to facilitate the above studies by providing the
community with tabulations of coefficients for the limb-darkening laws
that match the intensity distribution of state-of-the-art atmosphere
models with the highest fidelity.

With that in mind, we begin in Sect.~\ref{sec:advances} by mentioning
a few of the recent observational results concerning exoplanets using
the above-mentioned instruments, where limb-darkening has played a
role. We proceed then to Sect.~\ref{sec:models}, where we describe the
stellar atmosphere models used to generate the LDCs, as well as the
numerical method we applied. LDCs are presented here for some of the
most used passbands of NIRCam, NIRISS, and NIRSpec. We end with
Sect.~\ref{sec:remarks}, in which we comment briefly on the quadratic
limb-darkening law --- one of the more popular ones in the exoplanet
field --- and its limitations.

\section{Advances in  transmission spectroscopy  and light curve analyses  using JWST} 
\label{sec:advances}

A large number of papers making use of the instruments installed on
board the JWST have appeared in the recent literature, and many more
are sure to come. In the very active field of exoplanets, some of
those studies have dealt with the analysis of transit light curves or
phase curves, and others with transmission spectroscopy. In both cases
they rely on limb-darkening laws, in one way or another.  To illustrate
some of that work, below is a list a few representative examples of
such studies, along with the key results they have obtained:

\noindent $\bullet$ Rustamkulov et al.\ (2023) analyzed the exoplanet
WASP-39b, of similar mass as Saturn, using the PRISM mode on JWST's
NIRSpec instrument. They reported the detection of the following
chemical species in the atmosphere of the planet: Na (19$\sigma$),
H$_2$O (33$\sigma$), CO$_2$ (28$\sigma$) and CO (7$\sigma$).

\noindent $\bullet$ Alderson et al.\ (2024) analyzed the super-Earth
TOI-836b in the G395H passband of the NIRSpec instrument, and
concluded that TOI-836b does not have an H$_2$ dominated atmosphere.

\noindent $\bullet$ Ahrer et al.\ (2023) also investigated the
exoplanet WASP-39b using NIRSpec/PRISM, and identified carbon dioxide
(CO$_2$) in its atmosphere. The same study detected water vapor, and
placed a limit on the abundance of methane.  They also concluded that
the atmospheric metallicity is between 1 and 100 times solar.

\noindent $\bullet$ Liu et al.\ (2025) studied the detectability of
planetary oblateness, spin, and obliquity, from light curves obtained with
NIRSpec/PRISM on JWST, using the exoplanet Kepler-167b as an
example. They showed that observations of a single transit would be
able to detect a Saturn-like oblateness ($f = 0.1$) with high
confidence, or set a stringent upper limit on the oblateness
parameter, so long as the planetary spin is slightly misaligned ($\geq
20^{\circ}$) with respect to the orbital axis.

\noindent $\bullet$ Feinstein et al.\ (2023) reported an observation
of the transmission spectrum of WASP-39b using the Single-Object
Slitless Spectroscopy (SOSS) mode of the Near Infrared Imager. From
their analysis of the atmosphere, they estimated the metallicity to be
in the range of 10--30 times solar.  They found a sub-solar
carbon-to-oxygen ratio, and a solar-to-super-solar potassium-to-oxygen
ratio.

\noindent $\bullet$ Lustig-Yaeger et al.\ (2023) used NIRSpec in the
G395H passband to validate and study the Earth-sized exoplanet
LHS~475b. They ruled out primordial hydrogen-dominated and cloudless
pure methane atmospheres.

\noindent $\bullet$ Wallack et al.\ (2024) investigated the spectrum
of the sub-Neptune TOI-836c using the G395H passband of NIRSpec.  The
authors ruled out atmospheric metallicities below 175 times solar in
the absence of aerosols at $\leq$ 1 millibar.

The above investigations, and many others like them that take
advantage of the observational capabilities of the JWST mission,
provide the motivation for supplying users with new LDCs for the
4-term and power-2 laws, covering 11 passbands used with the NIRCam,
NIRISS, and NIRSpec instruments. The limitations of other
limb-darkening parametrizations have been pointed out in some of the
above studies (e.g., Ahrer et al. 2023), and we expect the new
coefficients calculated here to bring improvements in the analysis of
observational data from this space mission.

\section{Stellar atmosphere models and methodology}
\label{sec:models}

In this paper we make use of state-of-the-art spherically symmetric
models based on the PHOENIX code (NewEra model grid;
  Hauschildt et al.\ 2025).  A 1D spherically symmetric model may be
described loosely as being made up of two parts: a core, and an
envelope.  Roughly speaking, the core behaves approximately as
prescribed by a plane-parallel model.  The envelope behaves
differently, as it carries the spherical part. We define
quasi-spherical models as those that do not consider the full
intensity drop-off region. This concept is useful for situations in
which small intensities near the limb coming from spherical models
cannot be observationally detected, and/or if the effects of
sphericity are negligible.

Here we employ the Levenberg–Marquardt least-squares minimization
method (LML), adapted to the non-linear case to derive the LDCs for
these models. Before applying the LML to each passband, the specific
intensities for the photometric systems of interest were integrated
using the following equation:
\begin{equation}
  I_{a}(\mu) = ({hc})^{-1} {\int_{\lambda_1}^{\lambda_2} {
      I(\lambda,\mu) \lambda S(\lambda)_{a}
      d\,\lambda}\over\int_{\lambda_1}^{\lambda_2} { S(\lambda)_{a}~
      d\,\lambda}}.
\end{equation}	
In this expression $h$ is the Planck constant, $\lambda$ is the
wavelength, $c$ is the speed of light in vacuum, $I_{a}(\mu)$ is the
specific intensity in passband $a$, $I(\lambda,\mu)$ is the
monochromatic specific intensity, and $S(\lambda)_a$ is the response
function. Table~\ref{tab:passbands} lists the JWST passbands
considered for this work, and their mean wavelengths.\footnote{The
  corresponding transmission functions for these and other JWST
  passbands may be obtained at
  \url{https://jwst-docs.stsci.edu/jwst-exposure-time-calculator-overview/jwst-etc-pandeia-engine-tutorial/jwst-etc-instrument-throughputs#gsc.tab=0},
  and plots can be generated at \url{https://jwst.etc.stsci.edu/}}

\begin{table}
  \caption{NIRCam, NIRISS, and NIRSpec passbands. \label{tab:passbands}}
  \begin{flushleft}
    \centering
    \begin{tabular}{lc}
      \hline
      Instrument and Passband & $\lambda_a$ ($\mu$m) \\
      \hline
      NIRCam F210M             & 2.10 \\
      NIRCam F322W2            & 3.34 \\
      NIRCam F444W             & 4.44 \\
      NIRISS SOSS Order 1      & 1.80 \\
      NIRISS SOSS Order 2      & 0.80 \\
      NIRISS F277W             & 2.80 \\
      NIRSpec G235H/F170LP     & 2.44 \\
      NIRSpec G235M/F170LP     & 2.45 \\
      NIRSpec G395H/F290LP     & 4.04 \\
      NIRSpec G395M/F290LP     & 4.03 \\
      NIRSpec PRISM            & 3.65 \\
      \hline
      \hline
    \end{tabular}
  \end{flushleft}
\end{table}

The LDC calculations were performed for a total of 306 PHOENIX models
specifically computed for the present work, with each atmosphere model
occupying more than 2.0~GB of disc space. In order to improve the
input physics for the analysis of light curves and transmission
spectroscopy, the models were originally computed at very high
resolution, with approximately $1.08 \times 10^6$ wavelength points
covering the range 0.1--6.0 $\mu$m that is relevant for the JWST
passbands of NIRCam, NIRISS, and NIRSpec. All models were calculated
for the solar abundance and a fixed microturbulent velocity of $\xi
= 1.0~\kms$ (as per Hauschildt et al.\ 2025), over an interval in surface gravity $\log g$ between 3.0
to 5.5, and effective temperatures $T_{\rm eff}$ from 2400\,K to
7800\,K.  In each case, we interpolated the original grid of 127 $\mu$
locations to form 60,000 regularly spaced points, in order to better
sample the limb of the star and avoid numerical inaccuracies that
would otherwise occur when using LML.

The specific intensities from spherical model atmospheres show a much
more pronounced curvature near the limb than their plane-parallel
model counterparts. In fact, the specific intensity for spherical
models drops to zero. For this reason, it is more challenging to
obtain good fits in that region with some of the simpler
limb-darkening laws mentioned earlier. Only more flexible ones, such
as the 4-term law, are able to provide acceptable fits for the entire
intensity profile, although even for that prescription some of the
passbands yield less than optimal fits (see, e.g., Fig.~1 by Claret et
al.\ 2012).

In an earlier attempt to better describe the intensity distributions,
Claret \& Hauschildt (2003) introduced the concept of quasi-spherical
models. However, this simple approach only takes into account the
points in the intensity profile before the drop-off, down to an
arbitrarily chosen value of $\mu$ referred to as the critical point,
$\mu_{\rm crit}$, which is the smallest value used in fitting a given
limb-darkening law.  A year later, Wittkowski et al.\ (2004)
introduced a more sophisticated approach, whereby instead of relying
on an arbitrarily selected critical point, they defined it as the value
of $\mu$ corresponding to the maximum of the derivative of the
specific intensity with respect to $r$, where $r = \sqrt{1-\mu^2}$.
This point corresponds to a Rosseland mean optical depth $\tau_R
\approx 1.0$. The present PHOENIX models were computed with sufficient
$\mu$ sampling in the drop-off region to allow us to determine the
required derivative more precisely. Here we adopt the methodology of
Wittkowski et al.\ (2004), which produces steeper profiles and is
therefore more difficult to adjust, but provides a better
representation of the spherical part of the atmosphere models.
The merit function used as a measure of the quality of the fit for
each passband is
\begin{equation}
  {\chi^2} = \sum_{i=1}^{N} \left( {y_i - Y_i}\right)^2, 
\end{equation} 
where $y_i$ is the model intensity at point $i$, $Y_i$ is the fitted
function at the same point, and $N$ is the number of $\mu$ points.

The limb-darkening coefficients calculated for the 4-term law are
presented in Table~2 for all passbands, and those for the power-2 law
in Table~3. Both are available at the CDS.

As an illustration, Fig.~1 and Fig.~2 present comparisons of the
  fits to the intensity profile for the 4-term and the power-2 laws,
  for two of the NIRISS passbands. Plots for the other passbands in
  Table~1 are similar. In general, the fits with the 4-term law give
  smaller $\chi^2$ values than those with the power-2 law. In some
  cases the $\chi^2$ values from the 4-term law can be up to 100 times
  smaller, while in others they are similar.  A graphical comparison
  of the fit qualities between the 4-term, power-2, and quadratic laws
  is shown in Figs.~3--5.

\begin{figure}
  \includegraphics[width=0.5\textwidth]{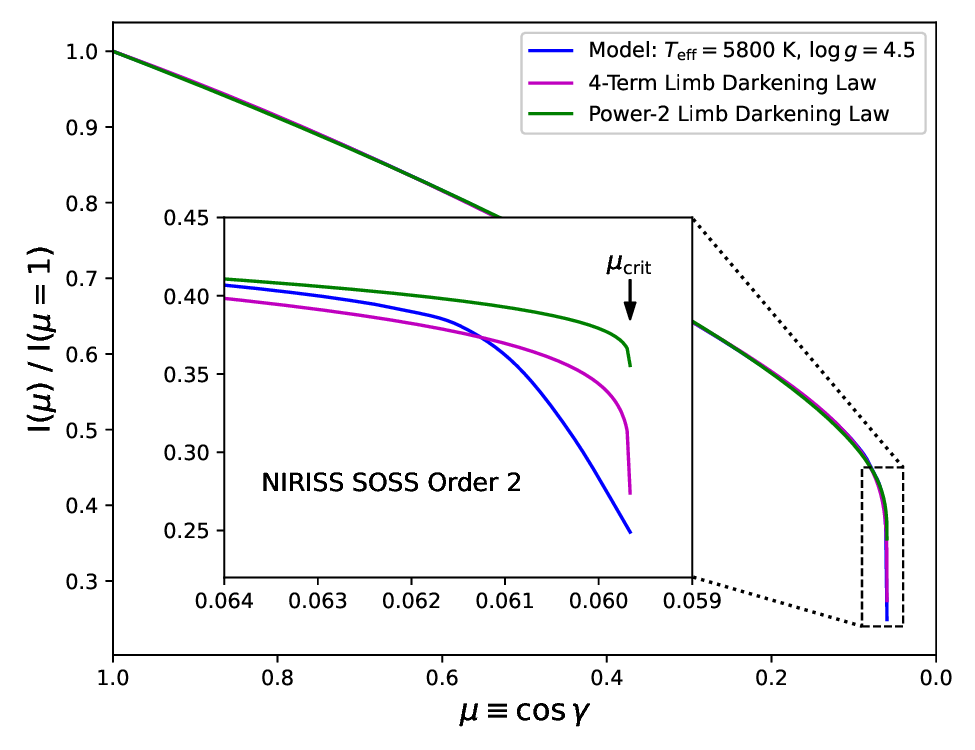}
  \caption{Angular distribution of the specific intensity for a
    model with $T_{\rm eff} = 5800$\,K, $\log g = 4.5$, ${\rm [M/H]} =
    0.0$, and $\xi = 1.0~\kms$ (blue line), for the
    NIRISS~SOSS~Order~2 passband. The fits provided by the 4-term and
    power-2 laws are displayed for comparison. The inset shows an
    enlargement around the drop-off region.}
\end{figure}

\begin{figure}
  \includegraphics[width=0.5\textwidth]{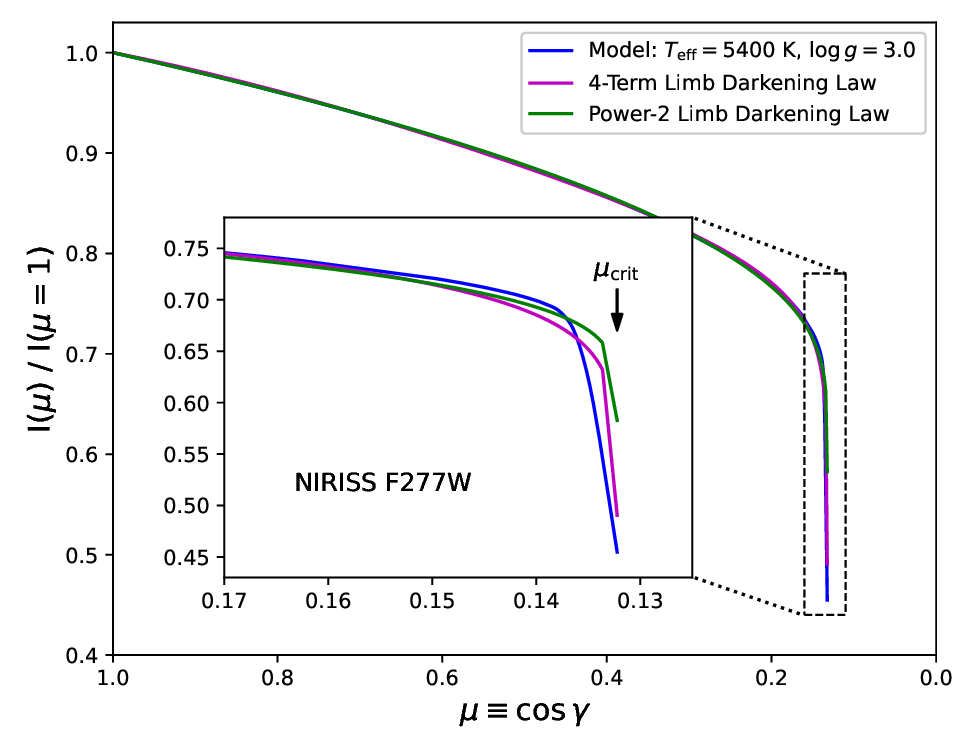}
  \caption{Similar to Fig.~1, for a model with $T_{\rm eff} =
    5400$\,K, $\log g = 3.0$, ${\rm [M/H]} = 0.0$, and $\xi =
    1.0~\kms$ (blue line). The NIRISS passband shown here is F277W.}
\end{figure}

\begin{figure}
  \includegraphics[width=0.5\textwidth]{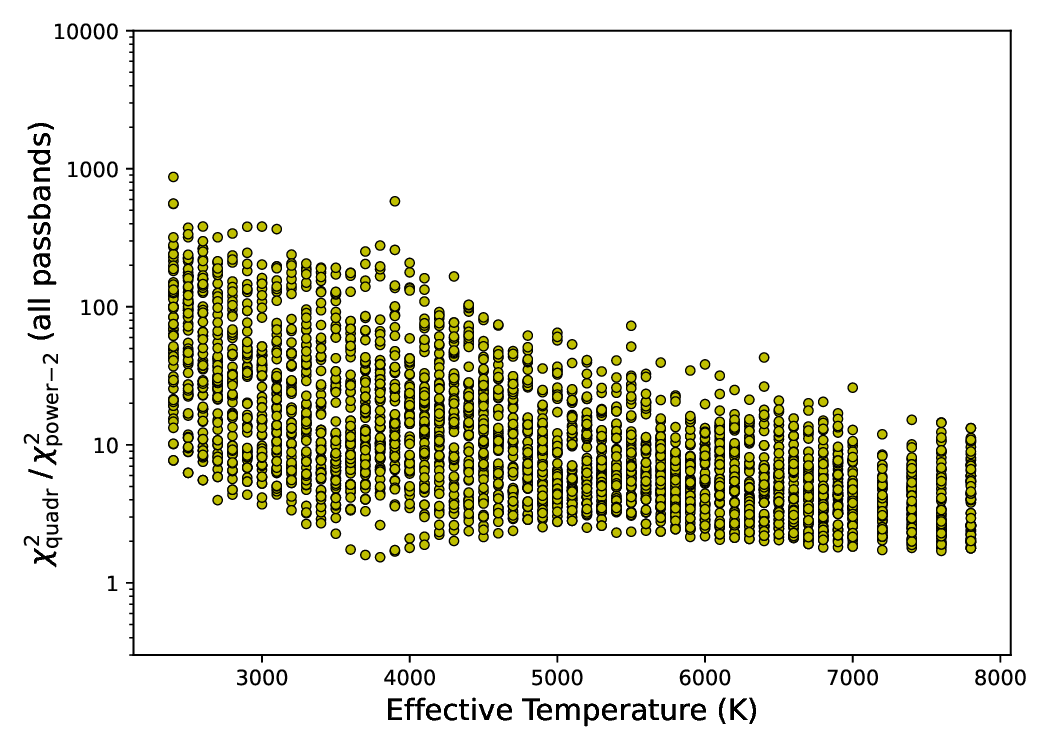}
  \caption{Comparison of the quality of the fits for all passbands
    studied here. The quality metric shown is the ratio between the
    $\chi^2$ values computed using the quadratic law and the power-2
    law, for all 306 models.}
\end{figure}

\begin{figure}
  \includegraphics[width=0.5\textwidth]{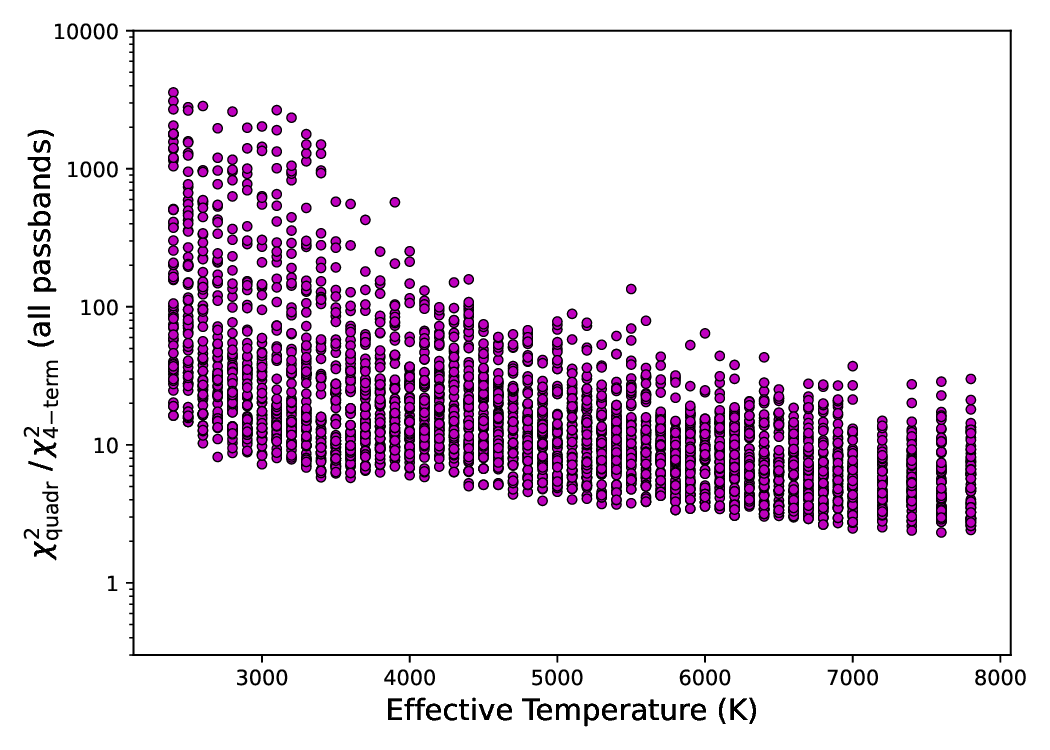}
  \caption{Similar to Fig.~3, for the ratio between the $\chi^2$
    values computed using the quadratic law and the 4-term law.}
\end{figure}

\begin{figure}
  \includegraphics[width=0.5\textwidth]{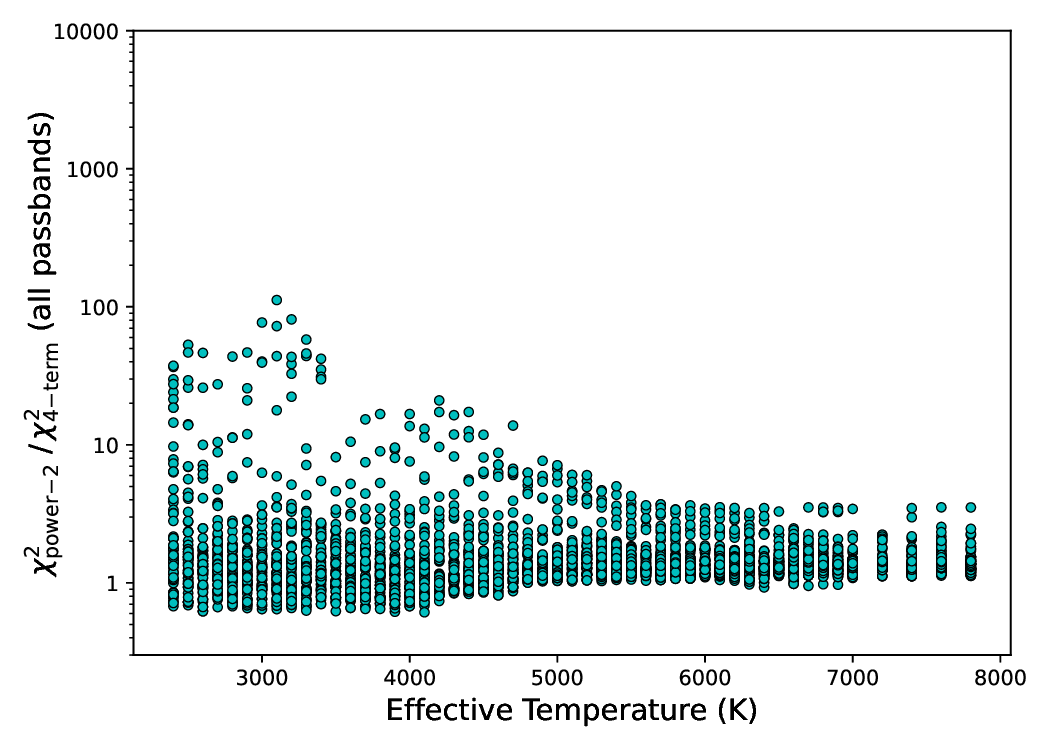}
  \caption{Similar to Fig.~3, for the ratio between the $\chi^2$
    values computed using the power-2 law and the 4-term law.}
\end{figure}

\section{Final remarks}
\label{sec:remarks}

Among the bi-parametric limb-darkening laws, the quadratic formula
(Eq.~\ref{eq:quadratic}) is one of the most commonly used in the
literature for the analysis of light curves and transmission
spectroscopy. We point out, however, that this prescription generally
does not provide accurate fits to the specific intensities produced by
realistic atmosphere models, for any given $T_{\rm eff}$, $\log g$, or
metallicity. This is the case even for models that adopt
plane-parallel geometry (Claret \& Southworth 2022; see their Figs.~2
and 3).  The $\chi^2$ values associated with such fits are quite high,
and the flux is not conserved within reasonable limits.  See
Figs.~3--5, where we compare the qualities of the fits among the
quadratic, power-2, and 4-term laws.  Use of the quadratic
prescription runs the risk of implicitly introducing into the analysis
a model of the stellar atmosphere that is not a good representation of
the one intended to begin with, characterized by certain values of
$T_{\rm eff}$, $\log g$, metallicity, and microturbulent velocity. As
a result, significant biases can be introduced into the analysis of
high-precision data such as JWST provides. Similar comments have been
made recently by Coulombe et al.\ (2024). In reality, of course, none
of the three laws mentioned previously (4-term, power-2, quadratic)
achieve a perfect fit to the distribution of specific intensities, as
they are only convenient approximations. However, based on the results
of this paper, and others discussing limb-darkening, it is advisable
to avoid using the quadratic law.  Instead, here we recommend the
4-term or power-2 laws, in that order of preference.  The LDCs
computed in this work for the 11 JWST passbands are available at the
CDS: Table~2 (4-term formula) and Table~3 (power-2
formula). Additional calculations can be carried out on request for
other photometric systems that are within the mentioned spectral range
of the present models.

\section{Data availability section}

Tables 2 and 3 are only available in electronic form at the CDS via
anonymous ftp to \url{cdsarc.u-strasbg.fr} (130.79.128.5) or via
\url{http://cdsweb.u-strasbg.fr/cgi-bin/qcat?J/A+A/}.

{}

\begin{acknowledgements}
  We thank the anonymous referee for helpful comments.  The Spanish
  MINC/AEI (PID2022-137241NB-C43 and PID2019-107061GB-C64) are
  gratefully acknowledged for their support during the development of
  this work. We also thank R.\ Morales and N.\ Robles (UDIT-IAA-CSIC)
  for their assistance with the calculations. AC also acknowledges
  financial support from grant CEX2021-001131-S, funded by
  MCIN/AEI/10.13039/501100011033.  The model grid calculations
  presented here were partially performed at the National Energy
  Research Supercomputer Center (NERSC), which is supported by the
  Office of Science of the U.S.\ Department of Energy under Contract
  No.\ DE-AC03-76SF00098. The authors gratefully acknowledge the
  computing time made available to them on the high-performance
  computers HLRN-IV at GWDG at the NHR Center NHR@G\"ottingen and at
  ZIB at the NHR Center NHR@Berlin. These Centers are jointly
  supported by the Federal Ministry of Education and Research, and the
  state governments participating in the NHR
  (www.nhr-verein.de/unsere-partner). Additional computing time was
  provided by the RRZ computing clusters Hummel and Hummel2. We thank
  all of these institutions for a generous allocation of computer
  time.  This research has made use of the SIMBAD database, operated
  at the CDS, Strasbourg, France, and of NASA's Astrophysics Data
  System Abstract Service.
\end{acknowledgements}

\end{document}